\documentclass{article}
\usepackage{epsfig, times} 
\newcommand{\beq}{\begin{equation}}
\newcommand{\eeq}{\end{equation}}
\newcommand{\beqa}{\begin{eqnarray}}
\newcommand{\eeqa}{\end{eqnarray}}

\begin{document}
\title{Is a neutron halo nothing but a large neutron skin? }
\author{Fumiyo Uchiyama   \\
University of Tsukuba\footnote{Former Professor}   
, 1-1-1 Tennodai, Tsukuba, Japan 305\footnote{
 Present mailing address: P.O.Box 14932, Berkeley 
         California, USA 94712}}
\maketitle
\begin{abstract}{We discuss the difference between neutron halos and neutron skins
matching to our usual thinking
of a skin as being connected to a body
whereas a halo is, in some sense,
disconnected from the body.
We emphasize that the existence of
one or more neutrons behaving incoherently with 
respect to the rest of the nucleons in a nucleus can match
the picture of
 a neutron halo. }

\end{abstract}


\section{Introduction }
In 1967, D. Wilkinson\cite{wilk}  suggested that in neutron rich nuclei,
a few neutron might be loosely bound forming "halos".
 Though it was unclear what 
precisely was meant by this terms, in the '80
the
unusual result  concerning 
the structures
 in longitudinal  momentum distribution \cite{ta}
 in $^{9}Li$ fragments from neutron rich $^{11}Li$ 
led physicists\cite{incoh1} to consider the  possibility of
 $^{11}Li$ being one of these new ``halo'' nuclei.
This is because, at the time,
    in high energy ($ \ge 1 Gev/c $)heavy ion experiments\cite{shroder} ,
 it was known that
momentum distributions of fragments produced peripherally
from
 ordinary stable
 nuclei such as 
$^{16}O$ in peripheral interactions were Gaussian whose widths
 were statistical averages of  Fermi motions of the 
 nucleons in the nuclei\cite{alf} (200-350 MeV/c)
while that of $^{11}Li$ was well fitted by the sum of two Gaussian
of a narrow width(~70 MeV/c) and an ordinary width.
In addition to the structures in momentum distribution,
 the unexpectedly large radius \cite{ta}  also indicated that
 $^{11}Li$ was an unusual nucleus (the number of 
neutrons in $^{11}Li$ is a magic number, 8 ).
Taking notice of
the small binding energy of two of the  neutrons in $^{11}Li$,
 the narrowness of one of the widths
   was  considered as the revelation of
 a ``halo'' structure \cite{incoh1} due to incoherent neutrons.
There have been many papers published  using 
 this loosely defined word, "halo"
:There have been theoretical analysis of halo nuclei  based on various models
as well as many experimental measurements for neutron and proton rich
  nuclei.   We refer to all these works \cite{many}
in the publications of proceedings of conferences 
and in the reference in review articles
for the related subjects in the reference except those directly concerned
with the questions relevant to my present work.

One usually think of a skin as being connected to a body 
on the surface, whereas a halo is in some sense disconnected or separated
from the body. How can we make this precise?
With such a vague definition, should
 $^{11}Li$ be considered  a halo nucleus?  How about $^{6}He$ ?
Isn't a ``halo'' nucleus a nucleus with just a large neutron skin?
If so, what is the size of radius or a  physical quantity dividing 
neutron rich nuclei 
into two categories, those with skin and the others with ``halo''? If ``halo''
does not have an implication of disconnectedness of some sort, wouldn't it 
be better to call it
``giant skin'' rather than  ``halo''?
Are a small binding energy of the last few neutrons 
 and-or a large radius
(cross sections) sufficient signatures for a nucleus to be described
as a new type of nucleus, a halo nucleus? 
These question arise because many neutron rich nuclei
(  $({N \over Z}) \geq  1 $ 
where N and Z are the neutron and charge numbers of a nucleus of nucleon number A=N+Z)
nuclei have neutron "skins" due to the fact that 
the last few neutrons occupy the large principal quantum number states
.  Therefore these last few neutrons
having smaller binding energy
compared to those of the other  nucleons in the core 
(which have binding energy of
 the order of Mev) 
can and thus make a neutron rich nuclei have 
 a larger cross sections comparing to that expected from
the number of nucleons, A.

Should we
call such nuclei halo? So where do we draw the line between 
neutron skin and neutron halo? Further more what happens to 
the notion that ``halo''
neutrons should be disconnected in some sense from the core? Can 
we find some qualitative criterion to distinguish halos from skin?  
The answer is yes. ``Halo'' neutrons should interact 
, in
disconnected way from the core,
 with impinging nuclei and particles.
It has been suggested\cite{incoh1}     that  neutrons in 
a ``halo'' behave 
  incoherently with respect to
 nucleons in the core in agreement with one's 
intuitive image of disconnectedness of a body(core), and its halo.

In this letter,
 we wish to clarify and explore this definition 
 in as  model independent a way as possible, namely phenomenologically.
I look for
physically measurable quantities which differ 
qualitatively from   nuclei 
 with skin.
Due to incoherence between core and the halo, the
observables such as cross sections
obtained from interaction with other particles and nuclei
are the sum of contributions from two parts, one from 
the core and the other from the halo. 
Upon interaction with impinging particles and nuclei the 
disconnected  core and halo will behave incoherently 
 with respect to each other.

There are many
successful models such as the Glauber model, the Skyrme method
,eikonal approximations, 
potential models, etc \cite{many}
which are  useful for relating different sets of data.
In my view,  however there is still no reliable low energy nuclear
theory 
 that can predict nuclear force saturation phenomena
such as halos  even
 though there exist new
 effective field theoretical attempts\cite{vector} that  
connect to QCD. For this reason,
we rely on  qualitative analysis.

\section{Neutron Skin or  Neutron Halo}
The difference between neutron skins and neutron halos has never been
analyzed explicitly \cite{mizu} and  the two terminologies are 
often  used interchangeably\cite{many}.
The loosely bound neutrons in the neutron distributions of some nuclei near drip line
form a  neutron skin. In the skin, the
 proton distribution is  small compared with
 that of neutron. The same is true for
neutron halos. Therefore the
 distinction between neutron skins
and neutron halos is difficult to make from  neutron distributions. 
Should we  call large neutron skins  neutron  halos? 
If that is the case,  what is the limit on the size of neutron skins 
above which they
should  be called 
 neutron halos\cite{mizu}? How do we take into account 
of the disconnectedness
of a halo from the body then? 
If the distinction is purely quantitative, should we just use a modifier, 
such as ``large'' or ``giant'',
to describe the neutron skin
instead of using an entirely different term, ``neutron halo''?.

The neutrons in 
halo or skin are bound 
 loosely regardless. Therefore the binding energy 
is not a good signature to phenomenologically distinguish the two. For a halo nucleus
to have a small binding energy and a large radius are necessary 
but not sufficient conditions.
As mentioned before, since one imagins a skin as being connected to a core 
whereas a halo is in some sense disconnected from the core,
we propose to distinguish halo neutrons  
from skin neutrons  
by the difference of their
 behaviors upon
interacting with the other nuclei or particles.
 We define halo neutrons to be those which
 behave incoherently with
respect to the nucleons in the core(the body  of a
halo nucleus)
while all nucleons
in ordinary  nuclei without halos  behave coherently as one unit 
 with strongly  interacting 
nuclei and particles in 
high energy peripheral  interactions. 
In other words, in a halo nucleus, there are two groups of nucleons
which behave  incoherently 
 with respect to each other.

Therefore for 
the measureable observables through the peripheral  interactions, 
incoherence of halo neutrons with respect to ordinary 
nucleons in the core appears as two independent 
components:  the probability for any physical process for a halo nucleus 
is the sum of two probabilities, the probability for the process to occur
with halo neutrons and the probability for the process to occur 
with the core nucleons.
Therefore, for instance,
the shapes of  differential cross section  of elementary 
particles 
scattered from ordinary 
nuclei  have the first diffraction
minimums at around the radii of the target nuclei in a diffractive energy region,
whereas
the differential cross sections from a halo nucleus
are  the sum of two differential cross sections each one of which have 
different diffraction minimums at different locations
due to the two disconnected
  nucleon distributions (which behave incoherently with respect to 
each other).
That is, in the case of a nucleus with a halo, what it means that neutrons in the halo 
act as  independent entities   distinguishable from
nucleons in the core, while for the case of neutrons in the 
skin, there is no independent, incoherent behavior
with respect to any other nucleons in the nucleus.

To express the above idea mathematically,
we express the wave functions for a  nucleus with halo  and a nucleus with skin
using one particle wave functions.
We denote the numbers of nucleons, protons, neutrons, and  neutrons in halo
by A, Z, N, and k, where A=Z+N, and k neutrons out of N are forming a 
halo around a core made out of Z protons and (N-k) neutrons.

The neutron distribution function for a nucleus having a
neutron skin is expressed by the square of the totally antisymmetrized 
wave function of N neutrons.
If we express the wave function in terms of
 single particle wave functions, $ \phi({\vec r})_{\alpha_{i}}$
for an example, the antisymmetrized wave function 
for a nucleus without halo can be written as a Slater determinant,

$$
\Psi_{neutron}(\vec r_{1}, \vec r_{2}, \ldots \vec r_{N}) = {1 \over \sqrt{N!}}\\
\left | \begin{array}{cccc}
             \phi_{\alpha_{1}}({\vec r_{1}})  & \phi_{\alpha_{2}}(\vec r_{1}) & \ldots & \phi_{\alpha_{N}}(\vec r_{1}) \\
             \phi_{\alpha_{1}}(\vec r_{2})  & \phi_{\alpha_{2}}(\vec r_{2}) & \ldots & \phi_{\alpha_{N}} (\vec r_{2}) \\
             \multicolumn{4}{c}{\dotfill} \\
             \multicolumn{4}{c}{\dotfill} \\
             \phi_{\alpha_{1}}(\vec r_{N})  &\phi_{\alpha_{2}}(\vec r_{N})  & \ldots & \phi_{\alpha_{N}}(\vec r_{N})
       \end{array}
  \right |  
$$
where $\alpha_{i}$ represent quantum numbers of the special wave functions.
 The so called neutron skin will come from the
 neutrons  
which are occupying  the states in the larger quantum numbers.
On the other hands, the wave function of  the halo nucleus 
made of Z protons and N neutrons out of which
 k neutrons are forming a halo, is
in good approximation 
expressed as
$$ 
\Psi_{neutron}(\vec r_{1}, \vec r_{2}, \ldots \vec r_{N}) = 
\Psi_{halo}(\vec r_{1}, \vec r_{2}, \ldots \vec r_{k}) 
\Psi^{n}_{core}(\vec r_{k+1}, \ldots \vec r_{N}) 
$$
where
$$
\Psi_{halo}(\vec {r_{1}}, \vec r_{2}, \ldots \vec r_{k}) =  {1 \over \sqrt{k!}}\\
\left | \begin{array}{cccc}
             \phi_{\beta_{1}}({\vec r_{1}})  & \phi_{\beta_{2}}(\vec r_{1}) & \dots & \phi_{\beta_{k}}(\vec r_{1}) \\
             \phi_{\beta_{1}}(\vec r_{2})  & \phi_{\beta_{2}}(\vec r_{2}) & \dots & \phi_{\beta_{k}}(\vec r_{2}) \\
             \multicolumn{4}{c}{\dotfill} \\
             \multicolumn{4}{c}{\dotfill} \\
             \phi_{\beta_{1}}(\vec r_{k})  &\phi_{\beta_{2}}(\vec r_{k})  & \dots & \phi_{\beta_{k}}(\vec r_{k})
       \end{array}
  \right |  
$$
and a similar expressions for (N-k) neutron for the wave function of the neutrons in the core,
$\Psi^{n}_{core}(\vec r_{k+1}, \ldots \vec r_{N}) $.
$$
\Psi^{n}_{core}(\vec r_{k+1}, \vec r_{k+2}, \ldots \vec r_{N}) = {1 \over \sqrt{(N-k)!}}\\
\left | \begin{array}{cccc}
             \phi_{\beta_{k+1}}({\vec r_{k+1}})  & \phi_{\beta_{k+2}}(\vec r_{k+1}) & \ldots & \phi_{\beta_{N}}(\vec r_{k+1}) \\
             \phi_{\beta_{k+1}}(\vec r_{k+2})  & \phi_{\beta_{k+2}}(\vec r_{k+2}) & \ldots & \phi_{\beta_{N}}  (\vec r_{k+2}) \\
             \multicolumn{4}{c}{\dotfill} \\
             \multicolumn{4}{c}{\dotfill} \\
             \phi_{\beta_{k+1}}(\vec r_{N})  &\phi_{\beta_{k+2}}(\vec r_{N})  & \ldots & \phi_{\beta_{N}}(\vec r_{N})
       \end{array}
  \right |  
$$
That is the neutron wave function of a halo nucleus, 
$\Psi_{neutron}$ is factored in 
two antisymmetrized wave functions; one for neutrons in the
halo and other for those
in the core.  Such factorized wave functions 
are frequently and implicitly used for a single neutron
halo nuclei in various theoretical papers\cite{many}\cite{tomp}.
Wave functions of a nucleus approximated and used
 in terms of the product of two separately antisymmetrized
wave functions in theoretical calculations and modelings
  all  belong to the latter, 
namely the wave functions for nuclei with halo nucleons.
 Here the neutron distribution in the outskirt
of ordinary neutron distribution come from the square of 
$ 
\Psi_{halo}(\vec r_{1}, \vec r_{2}, \ldots \vec r_{k}) 
$ 
 in the equation.

The phenomenological obvious differences of halo nuclei
 appear,  as 
superposition of two components of physical quantities such as differential
cross sections, distributions
 of  longitudinal momentum of fragments
from a halo nucleus, and others.
We discuss in the next section
the signatures of halos in momentum distribution of fragments.
 Halo structures produce two shifts of the centers
of the two components in the momentum distribution, one with a narrow width
\cite{ta,incoh1, mis}  and the other with an ordinary width.
For light halo nuclei, the latter 
may submerge into its  back ground due to the large
contribution from halo neutrons or vice versa, but
for heavy nuclei with halo, the both components of 
the momentum distributions
will become more pronounced.
One  new 
aspect  of  the halo structure we could observe is
the shifts of the centers of the symmetric 
(Gaussian for spherically symmetric
nuclei\cite{fnote}) momentum
distributions in the  rest system of the fragmenting nuclei due to 
different amounts of recoil.
The two different shifts produce the asymmetry
in the overall momentum distributions of fragments broken-up
from a halo nucleus via peripheral interactions.

\section{Recoil Momentum in  Fragments from  Nuclei with Halo}
Years ago , in high energy heavy ion fragmentation experiments, the
longitudinal  momentum distributions
of fragments in peripheral interactions were analyzed:
The reactions can be written  in general form including target  by 
\beq
     B_{Z} + D \rightarrow  C_{z'} +anything
\eeq
where B, C and D are the nucleon numbers of beam, 
the observed fragments produced by the {\underline{peripheral interaction}}
, and the target nuclei
 respectively.  Z and z' are the charge numbers of
 the beam and it's observed fragment nuclei respectively.

Greiner et.al \cite{dug} showed that  Gaussian shapes provide
good fits to the observed momentum spectra for all isotopes, $C_{z'}$,
produced by fragmentation of beam nuclei 
regardless of beam, energy, target except in the case of hydrogen isotopes.
Their fits showed  that the centers of Gaussian
distributions are shifted backward relative to direction of beam
 about several tens Mev. 
Their beam and target nuclei 
are all ordinary nuclei (core) such as $^{16}O$ and $^{12}C$ (
nuclear spin J=0)
and the beam energies used were
1.05 and 2.1 Gev/nucleon. The outcome of such an experiment
 will not differ 
even for lower energy beams as long as 
the beam energies are much larger than the binding energies of the beam 
and target nuclei.
The shifts of the center of momentum distribution
of fragments in the rest system of the beam nucleus 
are  understood as the recoil \cite{masuda} 
of the beam nucleus upon collisions with target. 
The difference in the amount of shift
for different isotopes reflects the internal structure, especially the binding energy
 of the beam nucleus.

When
 the beam nuclei $B_{z}$, are
halo  nuclei 
 and when the observed  isotopes
$C_{z'}$ in the final state have the same quantum numbers 
as those of the  core of the halo beam nuclei,
the longitudinal momentum distribution of $C_{z'}$
is expected to be a superposition of two symmetric functions (it is a Gaussian
for $B_{Z}$ of  J=0)
around negatively shifted center points(we use a generic term ``Gaussian'' for this 
symmetric functions here after) respectively in the rest system of the beam:
Not only  the widths of the two ``Gaussian'' functions are  different, 
but the two ``Gaussian'' distributions are expected to  have different amount of 
shifts in  the centers 
of their Gaussian distributions in the rest system of
the beam depending on
how the $C_{z'}$ are produced. The relevant point is that there are
{\it{two components}} and 
is not the specific 
shapes of 
momentum distributions as the shapes depend on the structures of  $B_{z}$
even in pheriheral interactions.

Therefore, in principle
one can extract in principle two values of the shifts
( one of which 
could be very small and could be consistent with zero due to 
extra small binding energy) along with the two widths by careful fitting the data
using two Gaussian forms to the momentum distributions of fragments
in the rest system of beam,
though the latter distribution of wider width
is less easily observable for light nuclei
 in peripheral interaction regions.  
 From the fact that the binding energy of the neutrons in halo
are smaller than those in core in general
, we expect that
the excitation energies exchanged by peripheral interactions
differs between the two cases, which appear as the  differences in the recoil.
Intuitively, we can understand that the shift for the wider width 
Gaussian distribution
is larger than the shift in the narrow width ``Gaussian'' distribution because the recoil is 
smaller for smaller binding energy.
 The shift, i.e.. the recoil, of
 the narrow component of momentum distribution being produced by stripping loosely bound
neutrons
can be negligible while the momentum distribution 
of wider width has a shift of a finite size. This implies that there is an asymmetry
in momentum distributions of  the isotope(with the same quantum number as the core
of a halo nucleus) produced by peripheral 
collisions. This asymmetry is more easily observed 
experimentally than the two distinctive Gaussian
distributions with two different center shifts.

Let us estimate the shifts for the nuclei for which the excitation energy 
distribution can be expressed phenomenologically as  
\beq
\rho(E) =  {exp_(- {\beta \over E }) \over E^{2}}
\eeq
where $\beta$ is a constant of the order of binding energy.
The average energy, $\bar E $ transfered from the target
to the beam is given as 
\beq
\bar {E} = {{\int {E  \rho (E) dE}} \over  {\int {\rho (E) dE}}}
\eeq
 Since both core and halo could have their own internal structures 
, they make it difficult to estimate
the absolute values of recoils observable as the shifts from the excitation energy alone. 
The ratio of the two shifts of the center of ``Gaussian'' 
 momentum distributions  
 is less dependent 
on the details of the structures of core and of halo. 
Denoting the shifts of the center of  momentum distributions with a narrow
and an ordinary widths
by $p^{c}_{core}$ and $p^{c}_{halo}$ respectively, we get 
\beq
 { p^{c}_{core} \over  p^{c}_{halo}}  =  \sqrt{ \beta_{core} \over \beta_{halo}}
\eeq
As an  example, if we take, in rather arbitrary manners,
  the binding energy of halo neutrons 
$\beta_{halo} = 0.33MeV$ and the removal energy of the same number of neutrons
  in the core
as $\beta_{core}$ = 1.0MeV\cite{note3}, we get the ratio, 1.7: 
 The shift for narrow components of momentum
distribution is smaller than the one in wider momentum distribution as we expected.
 The situation is 
depicted in Fig. 1.
 
Observation of  the asymmetry due to the difference of the two shifts,
smaller shift in the narrow components of momentum distribution
comparing to that in wider momentum distributions,
gives another confirmation on the signature of existence of neutron halo.
The asymmetry is more exaggerated in the negative 
direction in the center of mass system of halo-nuclei.
The asymmetry in the over all momentum distribution of an
isotope fragment other than
the isotope with the same quantum numbers as those of core
will have less pronounced signatures of existence of neutron halo
due to the complexity of the processes.
 The asymmetry caused by the differences in shifts are not conspicuous
in the region of the peak
even though 
 the hight 
of the  ordinary distribution
is chosen to be 10 percents of that of the narrow distribution.
 However
a careful experimental data taking such as ref. (\cite{mis}) 
and  analysis using two 
``Gaussian'' forms with two center shifts( one of them can be zero) 
will enable us to even more strengthen the proof of 
the existence of halo structures
even though many other ``back ground''  processes due to the complex nuclear
structures other than the halo structure may obscure this signature.
\section{Summary and concluding remarks}
One usually thinks of a skin as being connected to a body 
whereas a halo is in some sense disconnected from the body.
Putting stress on to this different characteristics of the halo and the skin,
we made an emphasis to discriminate a nucleus with a neutron halo 
against a nucleus with an neutron skin by 
the existence of a few halo neutrons which behave 
incoherently with the nucleons in the core
of the nucleus.  
An experimental proof for the existance of incoherently behaving neutrons
are essential to determine a neutron rich nucleus to have 
a large skin or a halo. We have shown that one of the proofs is
to show there are two components in a physical quantity such as 
longitudinal as well as transverse momentum observed through
high energy {\underline{peripheral}} interactions 
and any small energy-momentum exchange between
beams and targets. 
The factorizability of
the wave functions
of halo nuclei stem out naturally from disconnectedness(incoherence).
Since the halo neutrons 
behave incoherently with nucleons in core, two independent 
components in observable are expected for a halo nucleus
and become the signatures of
halo.

\section{Acknowledgments}
The support from physics Department of University of California, Berkeley
is acknowledged. Author thanks Drs. N.
Masuda and M.Redlich
  for comments and Akop Pogosian for
computer assistance.

\newpage
\pagestyle{empty}
\begin{figure}[ht]
\begin{center}
\vspace*{0.0cm}
\epsfig{file=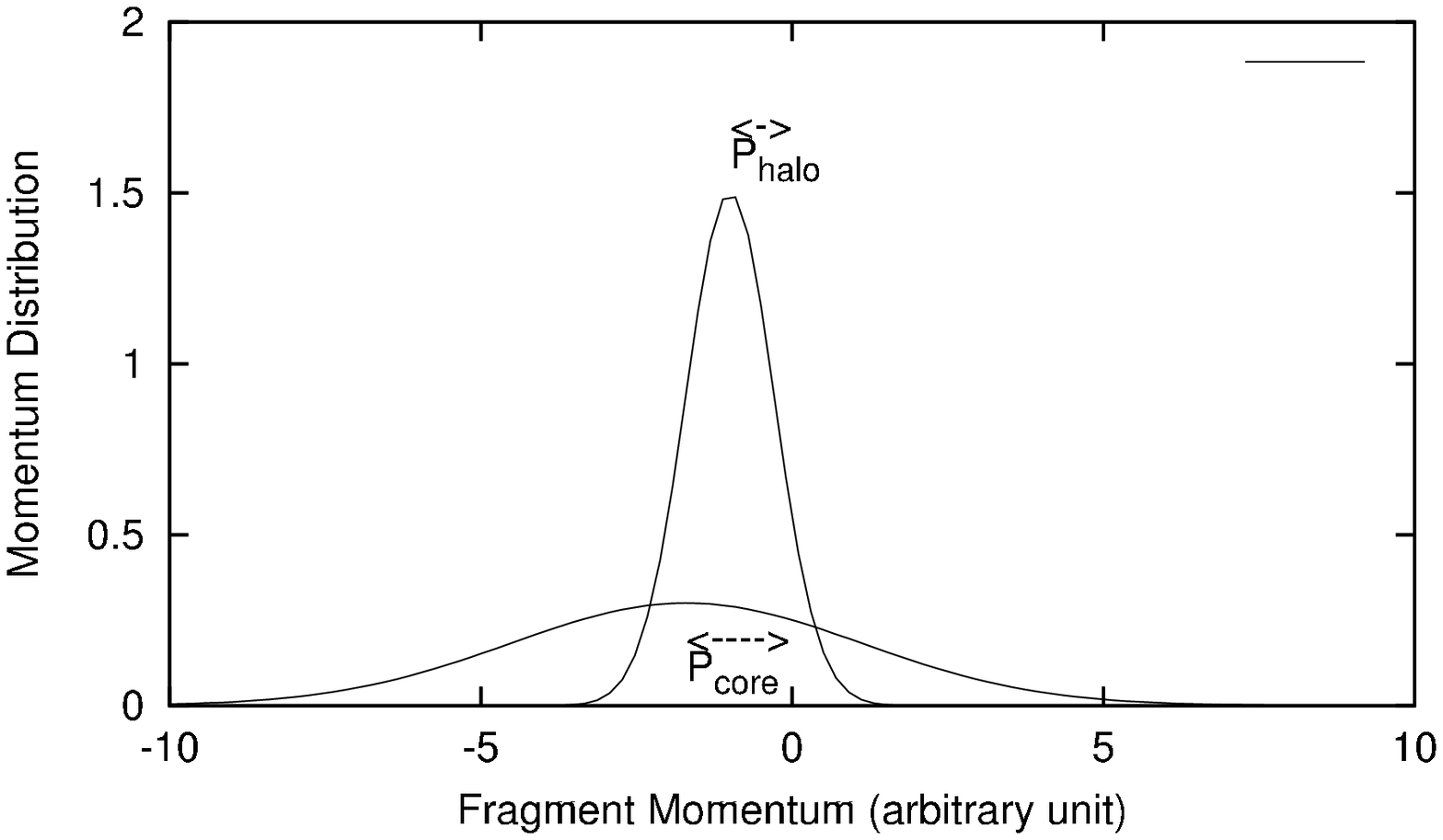, height=8.0cm}
\caption{ Schematic graph of the two components of
momentum distributions  of an isotope produced
by peripheral interactions. Summing the two contributions, the asymmetry in
the momentum distribution of fragments from a halo-nucleus are visible.
The magnitudes are arbitrary taken.}
\end{center}
\end{figure}


\begin{thebibliography}{}
\bibitem{wilk} D. Wilkinson, Comments Nuclear and  Particle  
 Physics 1, (1967) 80.
\bibitem{ta} T. Kobayashi el.al.,  Physical Review Letters  60,(1988)
2599. 
\bibitem{incoh1} F. Uchiyama and N. Masuda, Physical Review, C38, 
(1988)2670. 
\bibitem{shroder}  G.E. Greiner, D.Heckman, P. Lindstrom, Biser
Phys. Review Letters, (1974);
 2nd Heavy Ion Summer Study, LBL3675 (1974)
 ed. Lee Schroeder,
\bibitem{alf} A. Goldhaber, Physics. Letters 53B,(1974) 306. Talks at
 ``2nd High Energy Heavy Ion Summer Study'',  ed. Lee. S. Schroeder. 
      
\bibitem{many} References in Ann. Rev. Nucl and Part. Science 45,
 591(1995) by P. G. Hansen, A. S. Jensen and B. Jonson; references
  in Heavy elements and related phenomena PART IV, (1999) 893
  by W. Greiner and R. K. Gupta; P.G. Hansen and B. M. Sherill, 
Nuclear Physics A 693, (2001) 133 and references therein.
And references in R.Chatterjee, Physical Review C, 044604(2003)
\bibitem{mis} N. A. Orr et.al., Physical Review C 51, (1995) 3116.
\bibitem{tos} J. A. Tostevin et.al, Physical Review 66C,(2002)024607. 
(2003). 
\bibitem{vector} H. Georgi, Physical Review Letters, 63 (1986):
 S. Weinberg, Physics letters, 70A, (1990) 1917:
 The references in D. T. Son and M. A. Stephanov, arXive:hep-ph/0304182.
\bibitem{mizu} S. Mizutori et.al., Physical Review C 61,(2001) 044326. This paper
asked the same question explicitly concluding the difference to a quantitative one.  
\bibitem{tomp} I.J.Tompson and Y. Suzuki, Nuclear Physics A. 693,(2001) 421. 
The paper has used factorized wave functions for their halo nuclei calculations.
\bibitem{fnote} 
The point made here is the existance of two components in the measurements.
 The specific symmetric shapes are not concern.
\bibitem{dug} D. E.Greiner et.al.,  Physical Review Letters 35, (1975) 152.
\bibitem{masuda} N. Masuda and F. Uchiyama, Physical Review C15, (1977) 1598.
\bibitem{note3} The numbers we used, $\beta_{halo} = 0.33MeV$ and 
 $\beta_{core}$ = 1.0MeV are shosen arbitrary;the point is that the stronger the binding 
the larger the shift.
\end{thebibliography}
\end{document}